\documentclass[graphicx,superscriptaddress,reprint,longbibliography]{revtex4-1}

\usepackage[version=3]{mhchem} 
\usepackage[T1]{fontenc}
\usepackage{siunitx}
\usepackage{graphicx}

\usepackage[normalem]{ulem}
\useunder{\uline}{\ul}{}
\begin{document}
\author{Robin J. Dolleman}
\email{R.J.Dolleman@tudelft.nl}
\affiliation{Kavli Institute of Nanoscience, Delft University of Technology, Lorentzweg 1, 2628 CJ, Delft, The Netherlands}
\author{Samer Houri}
\affiliation{Kavli Institute of Nanoscience, Delft University of Technology, Lorentzweg 1, 2628 CJ, Delft, The Netherlands}
\affiliation{NTT Basic Research Laboratories, NTT Corporation, 3-1, Morinosato Wakamiya, Atsugi, Kanagawa, 243-0198, Japan}
\author{Abhilash Chandrashekar}
\affiliation{Department of Precision and Microsystems Engineering, Delft University of Technology, Mekelweg 2, 2628 CD, Delft, The Netherlands}
\author{Farbod Alijani}
\affiliation{Department of Precision and Microsystems Engineering, Delft University of Technology, Mekelweg 2, 2628 CD, Delft, The Netherlands}
\author{Herre S. J. van der Zant}
\affiliation{Kavli Institute of Nanoscience, Delft University of Technology, Lorentzweg 1, 2628 CJ, Delft, The Netherlands}
\author{Peter G. Steeneken}
\email{P.G.Steeneken@tudelft.nl}
\affiliation{Kavli Institute of Nanoscience, Delft University of Technology, Lorentzweg 1, 2628 CJ, Delft, The Netherlands}
\affiliation{Department of Precision and Microsystems Engineering, Delft University of Technology, Mekelweg 2, 2628 CD, Delft, The Netherlands}
\title{Graphene multi-mode parametric oscillators}

\begin{abstract}
In the field of nanomechanics, parametric excitations are of interest since they can greatly enhance sensing capabilities and eliminate cross-talk. However, parametric excitations often rely on externally tuned springs, which limits their application to high quality factor resonators and usually does not allow excitation of multiple higher modes into parametric resonance. Here we demonstrate parametric amplification and resonance of suspended single-layer graphene membranes by an efficient opto-thermal drive that modulates the intrinsic spring constant. With a large amplitude of the optical drive, a record number of 14 mechanical modes can be brought into parametric resonance by modulating a single parameter: the pretension. In contrast to conventional mechanical resonators, it is shown that graphene membranes demonstrate an interesting combination of both strong nonlinear stiffness and nonlinear damping. 
\end{abstract}
\maketitle

The history of parametric oscillations dates back to the 19th century and the observation of surface waves in the famous singing wineglass experiment of Michael Faraday \cite{faraday}. The advent of micro and nano engineering brought to life new ideas for exploiting parametric excitation for enhancing force and mass sensitivity \cite{turner1998five,rugar1991mechanical,karabalin2009parametric,karabalin2011signal,zhang2005application,zhang2004mass,zhang2002effect}, effective quality factor \cite{mahboob2008piezoelectrically}, and signal to noise ratio \cite{rugar1991mechanical} of tiny resonators. To date, many sensors, including gyroscopes \cite{oropeza2005parametric,hu2011parametrically,harish2008experimental}, mass sensors \cite{zhang2005application,zhang2004mass,zhang2002effect} and even mechanical memories \cite{mahboob2008bit,mahboob2014multimode,roukes2004mechanical,freeman2008nems} employ parametric excitation for improved performance.

Parametric micromechanical oscillators normally rely on stiffness modulation. The mechanical stiffness of microbeams is however fully determined by their fixed Young's modulus and geometry; parametric excitation can therefore usually only be achieved by modulating the stiffness of an externally applied spring, for example a voltage controlled electrostatic spring \cite{turner1998five,rugar1991mechanical,carr2000parametric,rhoads2010nonlinear,mathew2016dynamical,C7NR05721K}. However, this method has a small modulation amplitude, limiting its application to resonators with a high quality factor. Besides, the external spring force usually cannot excite all the higher mechanical modes of a single mechanical element. 

Here, we overcome these limitations by using tension modulation for parametric driving of suspended graphene membranes. Since graphene membranes have negligible bending rigidity, their stiffness is dominated by their pretension that can be efficiently modulated by heat \cite{ye2017very,yang2017local,davidovikj2017thermal}. The tension modulation $\Delta n_0(t)$ is given by $\Delta n_0 = \alpha E_{\mathrm{2D}} \Delta T$, where $\alpha$ is the thermal expansion coefficient, $E_{\mathrm{2D}}$ the 2D Young's modulus and $\Delta T$ the temperature modulation. Using approximate values from literature \cite{lee2008measurement,yoon2011negative}, one finds that $\Delta n_0(t) \approx 0.003 \Delta T$ Nm$^{-1}$K$^{-1}$, which means that a temperature modulation of 1 K already results in a tension modulation of the order of the intrinsic pretension $n_0$ (estimated to be between 0.003 N/m and 0.03 N/m \cite{dolleman2017optomechanics}) of the graphene membranes studied here. One can define the relative shift of the resonance frequency per unit of temperature as a figure of merit for the efficiency of the opto-thermal parametric drive: $\frac{1}{f_{\mathrm{res}}} \frac{\Delta f_{\mathrm{res}}}{\Delta T} = 0.1$ to 1 K$^{-1}$. This estimated value for graphene is 500-5000 times larger than in other optically pumped oscillators \cite{zalalutdinov2001optically}, thus illustrating that the parametric driving scheme for graphene membranes is possibly the most efficient method for reaching parametric oscillation in mechanical systems. Another advantage of tension modulation in membranes is that all mechanical degrees of freedom can be parametrically excited, since the effective stiffness of every mechanical mode of the system is proportional to $n_0$.  

We demonstrate the opto-thermal tension modulation technique experimentally on single layer graphene membranes and show that as many as 14 mechanical modes can be parametrically excited. A detailed analysis of both the directly and parametrically driven response of the fundamental mode reveals that nonlinear damping is essential to describe the mechanical motion. We experimentally demonstrate parametric signal amplification in these resonators, raising the amplitude of motion and increasing the effective quality factor of resonance. Finally, by analyzing the parametric resonances, we obtain information on the dissipation mechanism and loss tangent in graphene membranes, a subject that has recently received much attention \cite{guttinger2017energy,eichler2011nonlinear,singh2016negative,croy2012nonlinear}.

\begin{figure*}[ht!]
\includegraphics{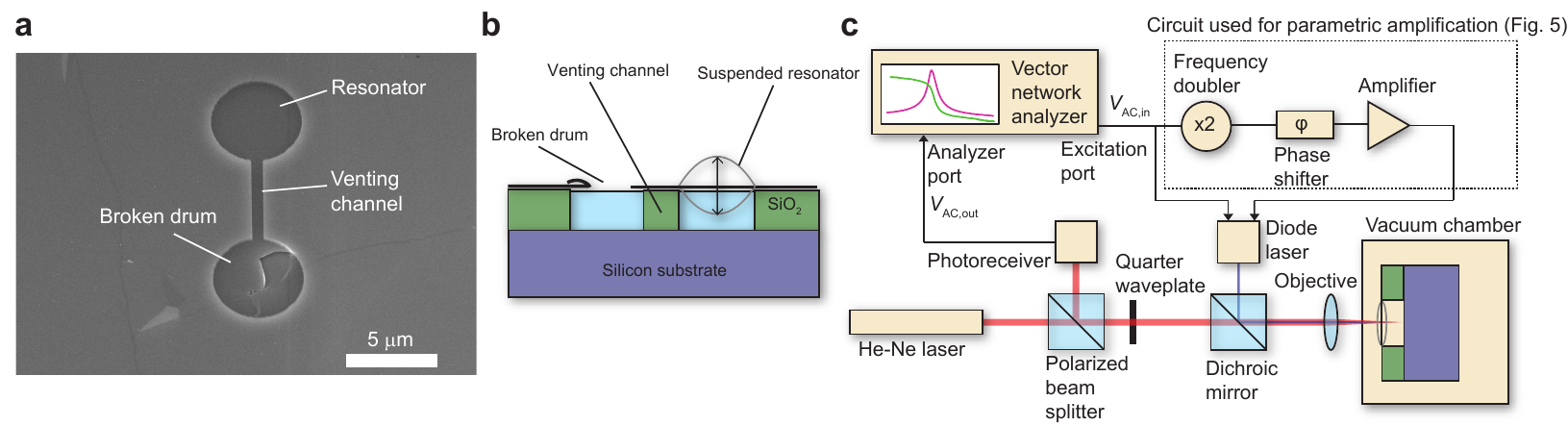}
\caption{\textbf{Single layer graphene resonators and the experimental setup.} \textbf{a}, Single layer graphene resonator under a scanning electron microscope (SEM). \textbf{b}, Cross section of the device (not to scale). \textbf{c}, Schematic of the measurement setup to actuate the membrane thermally and detect its motion by interferometry. \label{fig:1}}
\end{figure*}

\section*{Opto-thermal parametric driving of graphene}
Experiments are performed on single-layer CVD graphene drum resonators with a diameter of 5 $\mu$m and a cavity depth of 300 nm. The drums have venting channels to the environment to prevent the trapping of gas in the cavity (Fig. \ref{fig:1}\textbf{a},\textbf{b}, see Methods section for details on the fabrication). To achieve parametric drive, we use the experimental setup shown in Fig. \ref{fig:1}\textbf{c}. The light from a blue diode laser is focused on the membrane and its intensity is modulated by an input voltage $V_{\mathrm{ac,in}}$. This periodically heats up the membrane and creates a parametric drive due to the thermal strain. Parametric resonance occurs if the parametric driving term $\delta$ exceeds a threshold $\delta_t = \frac{2\omega_0^2}{Q}$, determined by the resonance frequency $\omega_0$ and quality factor $Q$ of resonance \cite{rugar1991mechanical}. At the same time, imperfections such as initial out-of-plane deformations, wrinkles and ripples in the membrane geometry enable this laser to directly drive the resonator by thermal expansion force, because thermal expansion will enhance these deformations and thus actuate the membrane. A more detailed discussion on this mechanism can be found in the Supplementary Information S3.

 A red helium-neon laser is used to read out the motion by the optical interference between the graphene membrane and the fixed substrate \cite{bunch2007electromechanical,castellanos2013single}. The ratio of the AC voltage amplitude generated by the photodetector and the AC driving voltage of the blue laser $V_{\mathrm{ac,out}}$/$V_{\mathrm{ac,in}}$ is determined by a vector network analyzer (VNA). The VNA has the ability to perform frequency conversion measurements, hence both homodyne and heterodyne detection can be performed in this setup such that both direct and parametric resonances can be analyzed.

\section*{Graphene multi-mode parametric oscillators}\label{sec:param}
 \begin{figure*}[t!]
\includegraphics{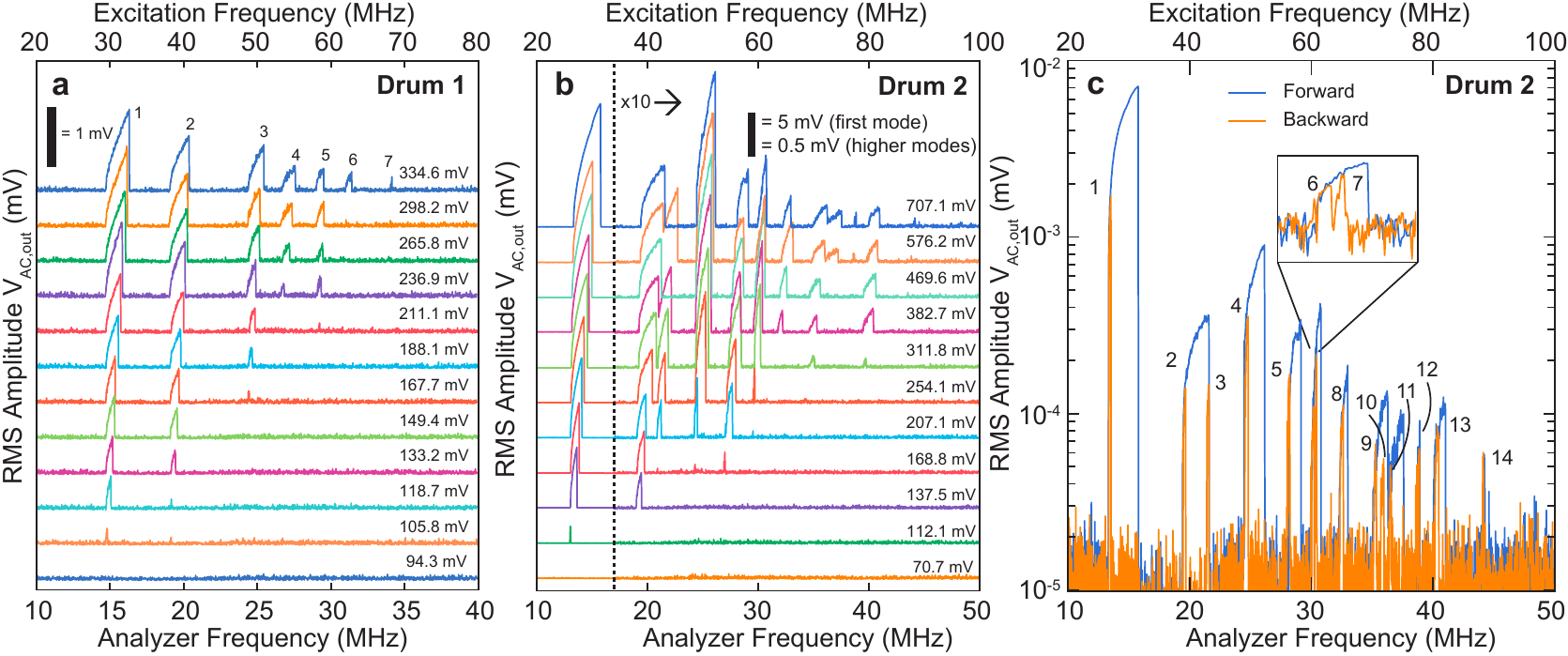}
\caption{\textbf{Multi-mode reponse of a parametrically driven graphene resonators.} \textbf{a} Waterfall plot of the multimode response at different driving amplitudes. Each mode appears at different driving levels due to variations in quality factor and effective driving force between them. The scale bar indicates the root mean square value (RMS) of $V_{\mathrm{ac,out}}$ and the labels on the right indicate the RMS driving amplitude $V_{\mathrm{ac,in}}$. \textbf{b} Waterfall plot for a different drum, showing more mechanical modes and modal interactions. \label{fig:5} \textbf{c} Forward and backward frequency sweep at the highest parametric driving amplitude for the drum in Fig. \ref{fig:5}\textbf{b}, revealing 14 distinct mechanical modes in parametric resonance.}
\end{figure*}
In Fig. \ref{fig:5}\textbf{a}, the blue laser is driven at 2$f$, while detecting the photodiode signal at $f$. When increasing the blue laser driving voltage $V_{\mathrm{ac,in}}$ a remarkable effect is observed. One-by-one, the parametric resonances of graphene appear, up to 7 different modes. Each mode reaches resonance at a different threshold driving amplitude $V_{\mathrm{ac,in}}$, due to differences in quality factor and the frequency dependence of the parametric driving parameter $\delta$ \cite{dolleman2017optomechanics}. The experiment is repeated on a different drum in Fig. \ref{fig:5}\textbf{b}. Interestingly,  in this case overlap between parametric resonances is observed at high driving levels. When overlap occurs, a direct transition between the high-amplitude solution of two adjacent parametric resonances is observed, e.g. at $V_{\mathrm{ac,in}}=382.7$ mV (RMS) between the second and third resonance. Interestingly, in some cases also transitions between the high-amplitude and low-amplitude solutions are observed, e.g. at  $V_{\mathrm{ac,in}}= 489.6$ mV (RMS) between the same 2 modes. 
This pseudorandom process is attributed to a strong dependence of the basin of attractions of the parametric high-amplitude and low-amplitude solutions on the initial conditions \cite{sanchez1990prediction}. Hence, the amplitude can fall into two stable solutions: either the high amplitude solution of the third mode or the zero amplitude solution of the third mode which is also observed at higher driving amplitudes ($V_{\mathrm{ac,in}}= 576.2$ and $707.1$ mV (RMS)). 

 Due to the overlap of parametric resonances in this drum, some resonances are skipped and not all resonances are found by sweeping from low to high frequency. Instead, when sweeping the frequency backward as shown in Fig. \ref{fig:5}\textbf{c}, as many as 14 parametric resonances are observed in this system. To our knowledge, this is the largest number of parametrically excited modes in a single mechanical element, as previously only 7 modes could be excited in cryogenic environments \cite{mahboob2014multimode}. 

 \section*{Mechanical nonlinearities in graphene resonators}\label{sec:nonlin}
 \begin{figure}
\includegraphics{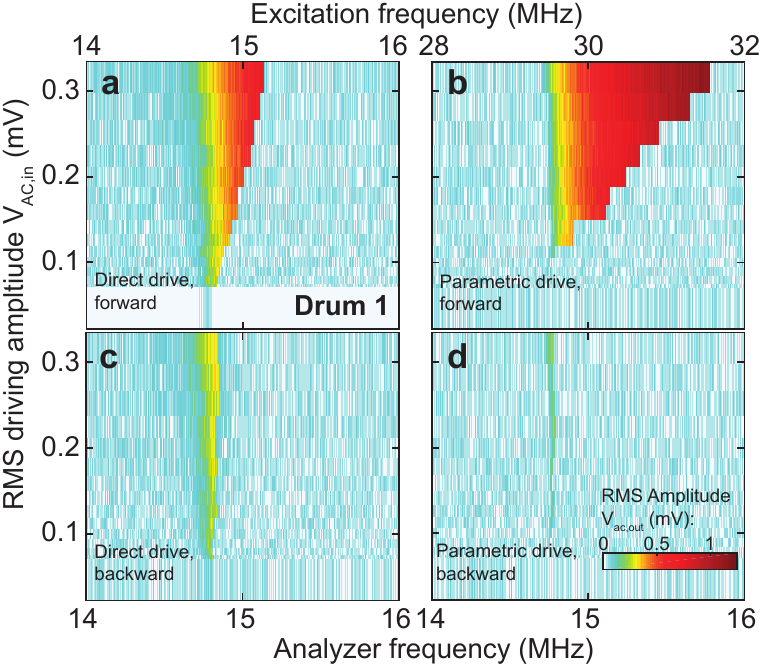}
\caption{\textbf{Frequency response of the fundamental mode to direct and parametric drive, for forward and backward frequency sweeps.} \textbf{a}, Direct drive with the frequency swept forwards. \textbf{b} Parametric drive with the frequency swept forwards. Below a driving threshold near $V_{\mathrm{ac,in}}\approx$ 0.11 mV (RMS) no mechanical response is observed. \textbf{c}, Direct drive with the frequency swept backwards. \textbf{d}, Parametric drive with the frequency swept backwards. \label{fig:2}}
\end{figure}
For a more detailed analysis of the physics, we focus on the frequency response of the fundamental mode to both direct and parametric drives. Figure \ref{fig:2} shows direct and parametric resonance of the fundamental mode as function of driving level, on a different drum than Fig. \ref{fig:5}. The VNA is configured to detect the directly driven frequency response (Fig. \ref{fig:2}\textbf{a},\textbf{c}). Sweeping the frequency forward (Fig. \ref{fig:2}\textbf{a}) and backward (Fig. \ref{fig:2}\textbf{b}) results in a hysteresis, that grows as the driving level is increased. This is typical for the geometric nonlinearity of the Duffing-type resonator, where the stiffness becomes larger at high amplitudes. In order to detect the parametric resonance, the VNA was configured in a heterodyne scheme at which $V_{\mathrm{ac,out}}$ is detected at half of the driving frequency $V_{\mathrm{ac,in}}$. Similar to the directly driven case, a hysteresis occurs between the forward (Fig. \ref{fig:2}\textbf{b}) and backward (Fig. \ref{fig:2}\textbf{d}) sweeps in frequency. Below an RMS drive amplitude of 0.11 mV, no response is observed. To show that the parametric resonance shows two stable phases of resonance separated by 180 degrees \cite{mahboob2008bit}, we performed an additional measurement which is shown in the Supplementary Information S2. 

 \begin{figure*}[t!]
\includegraphics{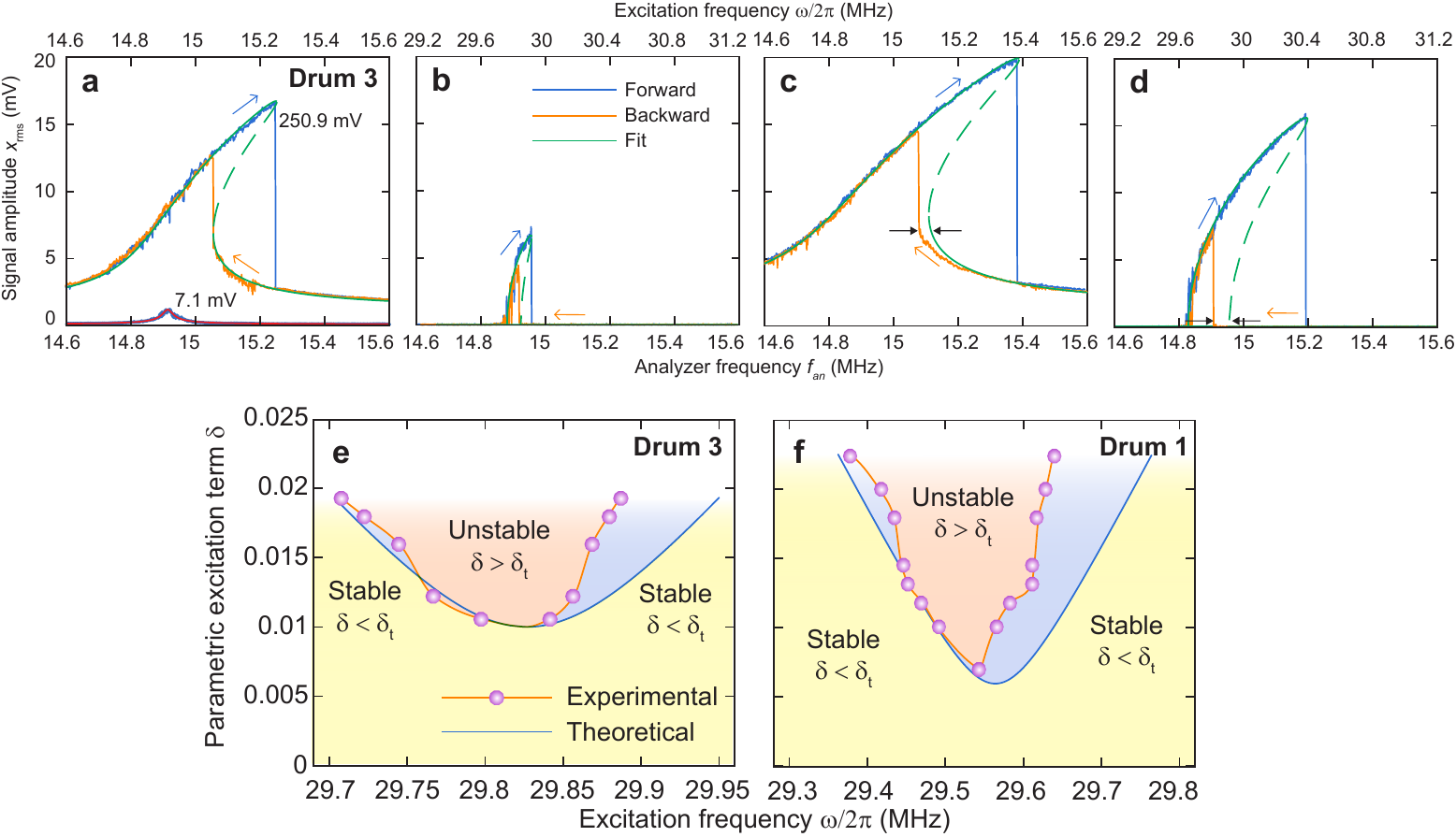}
\caption{\textbf{Comparison of experimental mechanical responses to theory.} \textbf{a}, Directly driven response at 7.1 and 250.9 mV RMS driving voltage and the fit obtained from Eq. \ref{eq:matthieudirect}. \textbf{b}, Parametric response and fit at 250.9 mV RMS driving voltage and the fit from Eq. \ref{eq:matthieudirect}. \textbf{c},  Directly driven response at 446.2 mV RMS driving level, the fit from Eq. \ref{eq:matthieudirect} shows a disagreement with the backward sweep, highlighted by black arrows. \textbf{d}, Parametric response at 446.2 mV (RMS). Black arrows highlight the disagreement between Eq. \ref{eq:matthieudirect} and experiment. \textbf{e}, Parametric resonance instability map for the fundamental mode of drum 2, compared to the prediction from Eq. \ref{eq:matthieudirect}. \textbf{f}, Parametric resonance instability map for the fundamental mode of drum 1 (Fig. \ref{fig:5}). \label{fig:8}}
\end{figure*}
Figure \ref{fig:8}\textbf{a}-\textbf{d} shows both directly and parametrically driven responses at different driving levels. In order to eludicate the effect of nonlinearities on the observed mechanical responses, a single degree-of-freedom model is derived that describes the motion of the resonator (see Supplementary Information S4) and this is fitted to the response curves in Fig. \ref{fig:8}\textbf{a}-\textbf{d} (see Supplementary Information S5). The model is a combination of the Duffing, van der Pol and Matthieu-Hill equation also used in other works \cite{eichler2011nonlinear,houri2017direct,lifshitz2008nonlinear,aubin2004limit}: 
\begin{equation}\label{eq:matthieudirect}
\ddot{x} + \mu \dot{x} + \nu x^2 \dot{x}+  (\beta  + \delta \cos{\omega t}) x + \gamma x^3 = F \cos{\omega t},
\end{equation} 
where $x$ is the displacement (which is approximately proportional to $V_{\mathrm{ac,out}}$), $\mu$ is the damping coefficient, $\nu$ the nonlinear damping coefficient, $\beta$ the linear stiffness coefficient, $\gamma$ the nonlinear stiffness coefficient, $\delta \cos{\omega t}$ the parametric driving and $F \cos{\omega t}$ the direct driving term. By setting $\gamma = 0$ and $\nu = 0$ one can fit the response at low drive level (Fig. \ref{fig:8}\textbf{a}) and obtain an initial value for $\mu$, $\beta$ and $F$. Initially fitting at high driving levels was attempted by setting $\nu = 0$, however it is found that such a model cannot account for the observed parametric response: nonlinear damping is indispensable to describe the maximum amplitude. Next, $\gamma$, $\nu$ and $F$ are used as fitting parameters to describe the nonlinear response (Fig. \ref{fig:8}\textbf{a}-\textbf{d}). Numerical values for the fit parameters are provided in the Supplementary information S1. 

 Figure \ref{fig:8} compares the fitted model and the experimental data for the directly- and parametrically driven fundamental resonance. This shows excellent agreement at lower driving levels (Fig. \ref{fig:8}\textbf{a}). We note that the fitting parameters $\mu$, $\nu$, $\beta$ and $\gamma$ are the same in the direct and parametric response within the error of the fitting procedure (see Supplementary information S1) and that both $\delta$ and $F$ are very nearly proportional to driving voltage $V_{\mathrm{AC,in}}$. It is however observed that the region of instability (Figs. \ref{fig:8}\textbf{e}, \textbf{f}) is narrower in our experiments than what is expected from eq. \ref{eq:matthieudirect}. 

\section*{Parametric signal amplification in graphene} \label{sec:amp}
Now we investigate the effects of parametric drive at low driving levels ($\delta < \delta_t$) by examining parametric amplification of the directly driven resonance. To measure parametric amplification, it is required to simultaneously drive the system at $f$ and $2 f$ (where $f$ is near the resonance frequency $f_0$). This is realized by splitting the driving circuit connected to the diode laser into two parts (Fig. \ref{fig:1}\textbf{c}). One path provides a small direct drive that excites the primary resonance of the membrane in the linear regime. The second path contains a frequency doubler, amplifier and phase shifter to enable parametric driving with controllable phase and gain with respect to the direct drive. A harmonic oscillator model is fitted to the response to extract the amplitude and the effective quality factor. The relation between amplitude gain $G$, parametric drive amplitude $\delta$ and phase shift $\phi$ of the direct drive is given by \cite{mahboob2008piezoelectrically,rugar1991mechanical}:
\begin{equation}\label{eq:gain}
G(\delta, \phi) = \left[ \frac{\cos^2\phi}{(1+\delta/\delta_t)^2} + \frac{\sin^2\phi}{(1-\delta/\delta_t)^2}\right]^{1/2}.
\end{equation}

\begin{figure}
\includegraphics{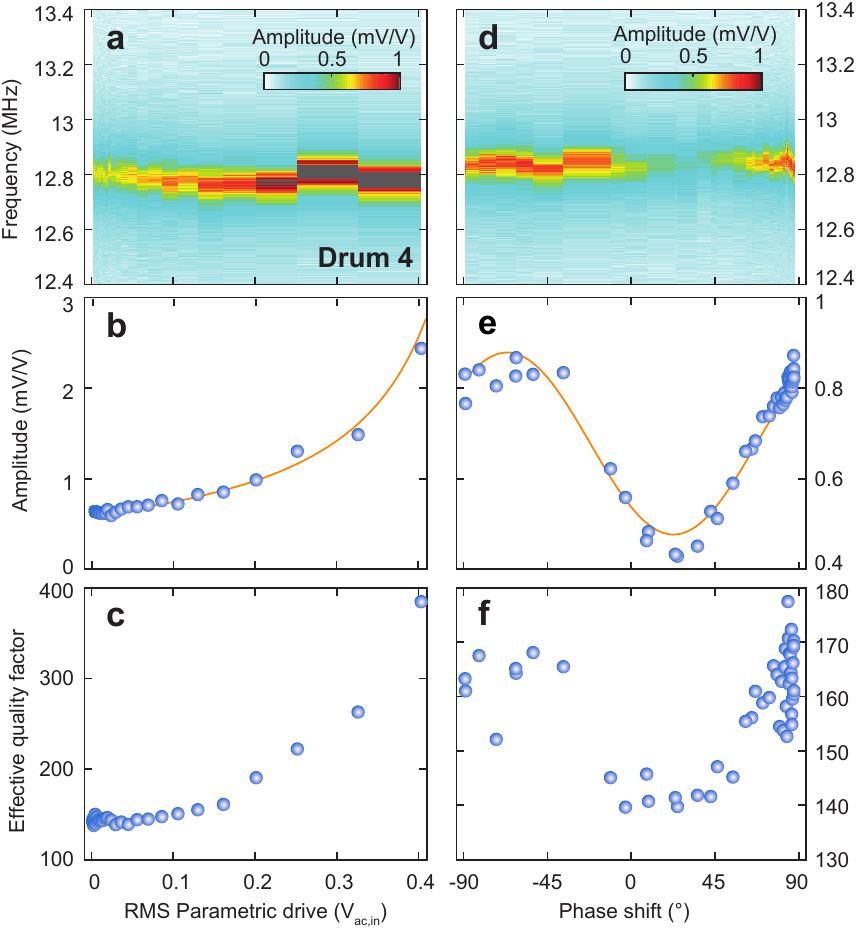}
\caption{\textbf{Parametric amplification in graphene: direct driven resonance with a sub-threshold ($\bf{\delta<\delta_t}$) parametric drive.} \textbf{a}, Transmission function of the direct drive as function of parametric drive. \textbf{b}, Amplitude of resonance obtained from a fit to a harmonic oscillator model as function of parametric drive, the red line is a fit to the theoretical behavior predicted by Eq. \ref{eq:gain}. \textbf{c}, Effective quality factor, obtained from a fit to a harmonic oscillator model, as function of parametric drive. \textbf{d}, Transmission funciton as  function of phase shift $\phi$. \textbf{e} Amplitude of resonance as function of phase $\phi$, the red line is a fit using eq. \ref{eq:gain}. \textbf{f}, Effective quality factor as function of phase $\phi$. \label{fig:6}}
\end{figure}
First, the amplification effect as function of parametric pumping amplitude in Fig. \ref{fig:6}\textbf{a} was examined by keeping the phase $\phi$ fixed at $\phi = $ -45 degrees. Increasing the amplitude of parametric drive increases the amplitude at resonance by a factor of 3-4 (Fig. \ref{fig:6}\textbf{b}) and the effective quality factor of resonance by almost a factor of 3 (Fig. \ref{fig:6}\textbf{c}). Figure \ref{fig:6}\textbf{d} shows that shifting the phase of the parametric drive significantly changes the amplitude of harmonic resonance. Figure \ref{fig:6}\textbf{e}-\textbf{f} shows that the gain $G$ and effective Q-factor $Q_{\mathrm{eff}}$ depend strongly on the phase of the parametric drive with respect to the direct drive. Fits of the data in Fig. \ref{fig:6}\textbf{b}, \textbf{e} show that the drive and phase-dependence of the parametric amplification is in accordance with theory. 

\section*{Discussion}
We have shown opto-thermal tension modulation as a mechanism to achieve large parametric excitations in graphene. The large parametric excitation enables low-$Q$ modes to be brought into parametric resonance. This allows parametric resonance study at room temperature (where graphene's $Q$-factor is a factor 100 lower than in the cryogenic regime \cite{chen2013graphene,zande2010large}), and also allows higher order modes (with lower Q) to be brought into parametric resonance. These beneficial conditions enable us to realize the graphene multi-mode parametric oscillators as shown in Fig. \ref{fig:5}. Due to their small mass and low spring constant, graphene membranes are very force sensitive. Here we demonstrate parametric amplification of graphene, which can further push graphene's limits of force sensitivity, since it enhances both the gain and effective quality factor of the resonators.

The asymmetry around $\omega_0$ observed in the region of instability (Fig  \ref{fig:8}) is a surprising result: such an asymmetry should not arise for the equation of motion (Eq. \ref{eq:matthieudirect}) used in the analysis. Something similar is observed in the directly driven response, where the lower saddle node bifurcation in the downward frequency sweep is always found at a lower frequency than simulated (Fig. \ref{fig:8}\textbf{c}) at high driving levels. Possibly, this indicates that the forcing terms are nonlinear \cite{rhoads2006generalized}. However, we find that both forcing terms $\delta$ and $F$ extracted from the fits are linear with the applied modulation amplitude and the forward frequency sweeps are well-described by this model (see Supplementary Information S1). The observed deviations (e.g. in Fig. \ref{fig:8}) can therefore not be explained by forcing nonlinearities.

The asymmetry and apparent decrease in resonance linewidth (Fig. \ref{fig:8}) thus suggest that a more unconventional dissipation model should be considered, including further terms to describe the amplitude-dependence of the dissipation. Similar deviations from conventional dissipation models have been previously found in multi-layered graphene resonators \cite{singh2016negative}, where it was concluded that the van-der-Pol term $\nu x^2 \dot{x}$ does not describe the nonlinear damping. Here we conclude that the van-der-Pol term is generally in agreement with the experiments, since it describes the saddle node bifurcation of the parametric resonances well, however additional dissipation terms might be needed to account for the asymmetry and narrowing of the parametric stability region (Fig. \ref{fig:8}\textbf{e},\textbf{f}).

The fit to the nonlinear response of the membrane allows us to extract a number for the Duffing ($\gamma$) and van-der-Pol terms ($\nu$) in our resonators. As shown in the Supplementary information S6, the mechanical loss tangent of graphene $\tan \delta_l$ at the resonance frequency can be determined from the ratio of these terms, $\tan \delta_l = \nu/\gamma$. From the values of the fits we obtain $\tan \delta_l = 0.34$ for drum 2 and $\tan \delta_l = 0.15$ for drum 3. The values of these loss tangents are in the same range as found by \citeauthor{jinkins2015examination}\cite{jinkins2015examination}. The obtained values for the loss tangent are relatively high for a crystalline material as graphene, therefore the observed nonlinear damping is likely not due to the intrinsic material properties but to other effects, such as sidewall adhesion  \cite{bunch2012adhesion} or unzipping of wrinkles \cite{ruiz2011softened}.

 Multi-mode parametric oscillators are interesting for applications where accurate frequency tracking of multiple modes is necessary. All of these modes can be parametrically amplified up to relatively high amplitudes. This is difficult to achieve with conventional oscillators that require a feedback loop and special filters or actuation schemes to make sure only the desired resonance is brought into oscillation. A possible application is inertial imaging \cite{hanay2015inertial}, where accurate tracking of multiple resonances allows one to determine the mass, location and shape of a particle on top of a resonator. In addition, these parametric oscillators can be used to build a binary information and computation system \cite{mahboob2014multimode}, where information is stored in the phase of the resonator. Multi-mode resonators have the potential of enabling parallel processing and data storage. The high resonance frequencies and relatively low Q of the graphene membranes can increase computation speed.  Moreover, since the driving frequency is double the readout frequency, parametric driving is less sensitive to cross-talk that is often present in directly driven resonators \cite{turner1998five} which can improve sensor performance.
 
 \section*{Conclusions}
In conclusion, we report on multi-mode parametric resonance and amplification in single layer graphene resonators by an opto-thermal tension modulation technique. It is demonstrated that the tension-dominated restoring force results in parametric excitation of multiple resonance modes in the system when the system is opto-thermally driven. The parametrically and directly driven resonances are compared to a single degree-of-freedom model based on the Duffing, van der Pol and Matthieu equations, with good agreement at low driving levels. This allows simultaneous determination of nonlinear stiffness and damping coefficients and results in a high-frequency determination of graphene's mechanical loss tangent. It was demonstrated that weak parametric drives can be used to amplify the motion and enhance the effective quality factor of resonance, that can potentially enhance the force sensitivity of graphene resonators. Graphene resonators are thus an interesting platform to study parametric excitations and their utilization for sensors with improved performance. 

\begin{acknowledgments}
We gratefully acknowledge Applied Nanolayers B.V. for the growth and transfer of the single layer graphene used in this study. We thank Y.M. Blanter, B.J. Gallacher, W.J. Venstra and S.W. Shaw for useful discussions and D. Davidovikj for help with scanning electron microscopy. This work is part of the research programme Integrated Graphene Pressure Sensors (IGPS) with project number 13307 which is financed by the Netherlands Organisation for Scientific Research (NWO).
The research leading to these results also received funding from the European Union's Horizon 2020 research and innovation programme under grant agreement No 649953 Graphene Flagship.
\end{acknowledgments}

\section*{Methods}
\subsection*{Fabrication}
Graphene resonators are fabricated by etching dumbbell-shaped cavities in a thermally grown, 285 nm SiO$_{2}$ layer on a silicon wafer. The etching did not fully stop at the silicon layer, resulting in cavities that are 300 nm thick. Circular membranes are formed by transfer of single layer chemical vapor deposited (CVD) graphene (Fig. \ref{fig:1}\textbf{a}). During the transfer process one side of the dumbbell is broken while the other side remains intact, creating a circular resonator on one side with a venting channel to the environment (Fig. \ref{fig:1}\textbf{b}). This prevents gas from being trapped in the cavity when the pressure in the surroundings changes. In the main section of this work four identical drums with a diameter of 5 micrometer are used; results obtained on drum 1 are shown in Figs. \ref{fig:5}\textbf{a}, \ref{fig:2} and \ref{fig:8}\textbf{f}, drum 2 in Figs. \ref{fig:5}\textbf{b},\textbf{c}, drum 3 in Figs. \ref{fig:8}\textbf{a}-\textbf{e} and drum 4 in Fig. \ref{fig:6}. A fifth drum was used in the Supplementary Information S2. More details on the fabrication and transfer process of the drum resonators can be found in ref. \cite{dolleman2017optomechanics}.

\subsection*{Experiments}
All measurements are performed at room temperature in a high vacuum environment with a pressure less than $\num{2e-5}$ mbar to minimize the effects of gas damping. The blue diode laser (Thorlabs LP405-SP10) has a wavelength of 405 nm and is biased with a 32 mA current, resulting in 0.76 mW of incident power measured before the objective. The red laser illuminates the sample with 1.2 mW of incident power (measured before the objective lens). The vector network analyzer is of type Rohde \& Schwarz ZNB4 with the frequency conversion option (k4) installed.

\section*{Author contributions}
R.J.D., S.H., F.A., A.C. and P.G.S. analyzed and interpreted the measurements. R.J.D. and S.H. performed the experiments. F.A., A.C. and P.G.S. derived the equation of motion and fits in Fig. \ref{fig:8}. H.S.J.v.d.Z. and P.G.S. supervised the project. The manuscript was jointly written by all authors with a main contribution from R.J.D.


\pagebreak
\onecolumngrid
\setcounter{equation}{0}
\setcounter{figure}{0}
\setcounter{table}{0}
\makeatletter
\renewcommand{\theequation}{S\arabic{equation}}
\renewcommand{\thefigure}{S\arabic{figure}}
\section*{Supplementary Information: Graphene multi-mode parametric oscillators}
In section S1, we show the complete dataset obtained while analyzing the nonlinearities of the graphene membrane. Section S2 shows an additional experiment that demonstrates the parametric oscillator has two stable phases. In section S3, an additional discussion is added that proposes why both direct and parametric excitations are observed in the same setup. Section S4 derives the equations of motion that has been used to perform the fitting and section S5 describes the numerical simulations used for the fitting procedure. Finally in section S6 the expression for the mechanical loss tangent of graphene is derived. 

\section*{S1: Complete datasets for the analysis of mechanical nonlinearities}
Figure 4 in the main text shows the fitting of the nonlinear mechanical response of the resonator (drum 3). In this section the remainder of this analysis is presented and the complete dataset from the fundamental mode of drum 1 is shown (Figs. 2\textbf{a}, 3, 4\textbf{f} in the main text).
\begin{figure}[h!]
\includegraphics{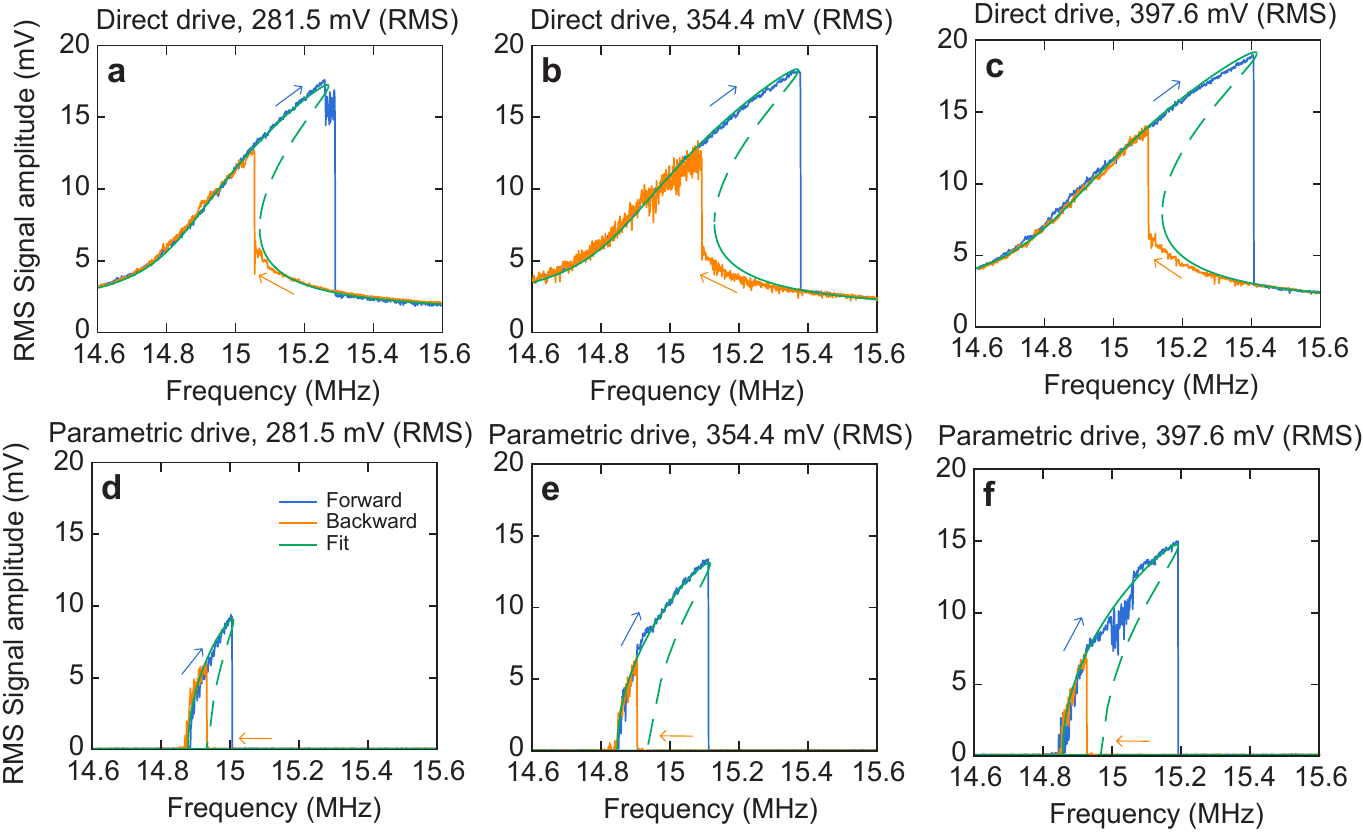}
\caption{Remainder of the dataset presented in Figs. 4 and 5 (a-e) in the main text. \label{fig:s1}}
\end{figure}

\begin{table}[h!]
\centering
\caption{Values obtained from the fits on the response from drum 3 in Fig. \ref{fig:s1} and Figs. 4\textbf{a}-\textbf{e} in the main text }
\label{tab:1l}
\begin{tabular}{llllllllll}
\hline
\hline\\
~ &    Direct drive    &    &       &                                                      & Parametric drive &        &    &       &                                                           \\\hline\\
RMS Drive (mV)       & $\mu$     &  $\nu$ & $\gamma$ & \begin{tabular}[c]{@{}l@{}}$F$ $\times$ 10$^{-5}$\end{tabular}           & $\mu$     & $\nu$ & $\gamma$ & \begin{tabular}[c]{@{}l@{}}$\delta$ $\times$ 10$^{-2}$\end{tabular} \\ \hline
250.9           & 0.0045 & 70 & 225   & 8                                                                  & 0.0045 & 76 & 215   & 1.06                                                      \\ 
281.5          & 0.0045 & 72 & 220   & 9.2                                                                & 0.0045 & 76 & 220   & 1.22                                                      \\ 
354.4           & 0.0045 & 74 & 230   & 11.7                                                             & 0.0045 & 79 & 225   & 1.6                                                       \\ 
397.6           & 0.0045 & 76 & 230   & 14.2                                                           & 0.0046 & 80 & 225   & 1.8                                                       \\
446.2           & 0.0045 & 76 & 230   & 15.5                                                             & 0.0046 & 80 & 225   & 1.93                                                      
\\ \hline \hline
\end{tabular}
\end{table}

\pagebreak
\subsection{Dataset of drum 1}
\begin{figure}[h!]
\includegraphics{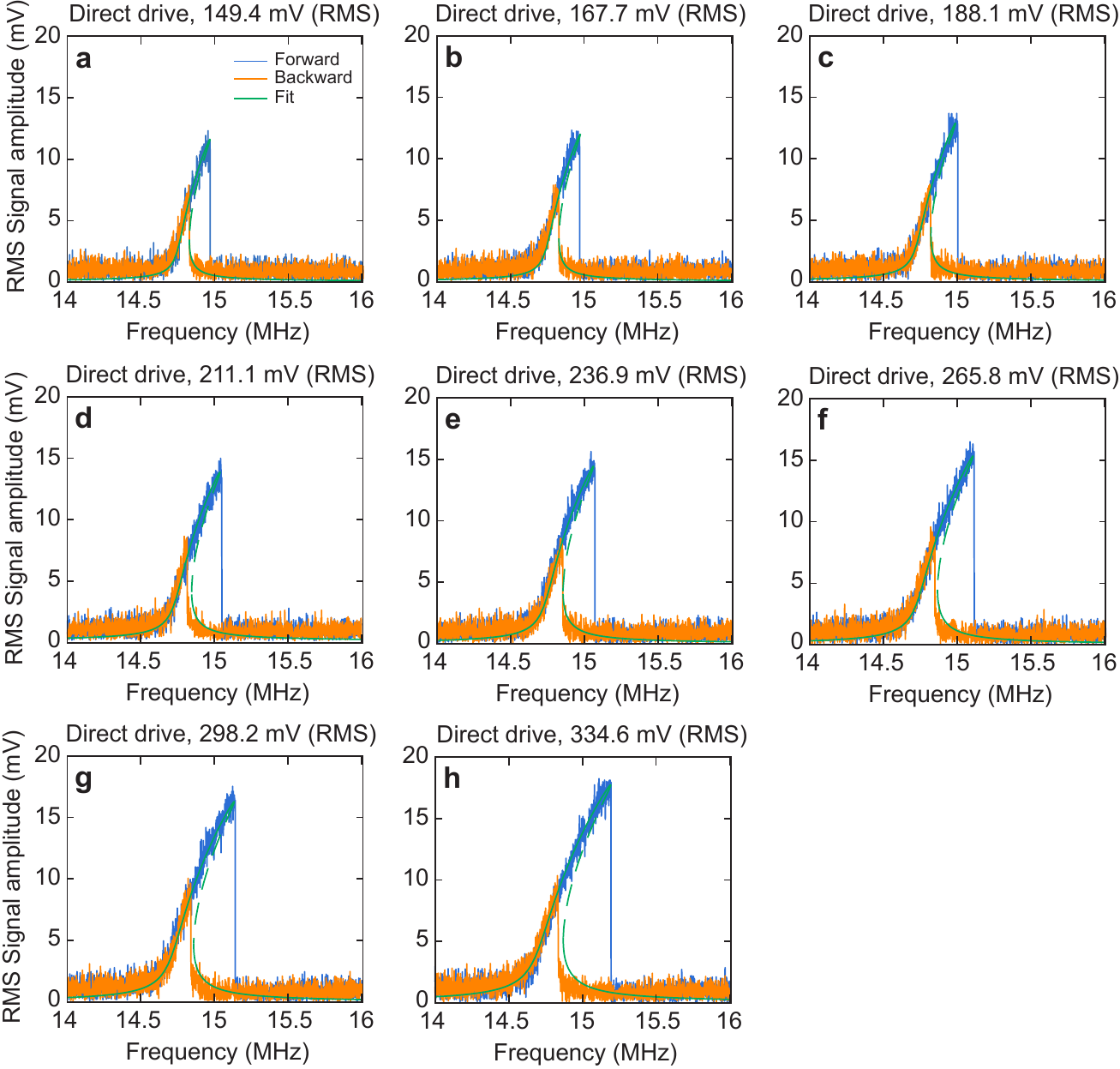}
\caption{Analysis of directly driven response of the fundamental mode of drum 1 (Figs. 2\textbf{a}, 3, 4\textbf{f} in the main text).  \label{fig:s2}}
\end{figure}
\begin{figure}[h!]
\includegraphics{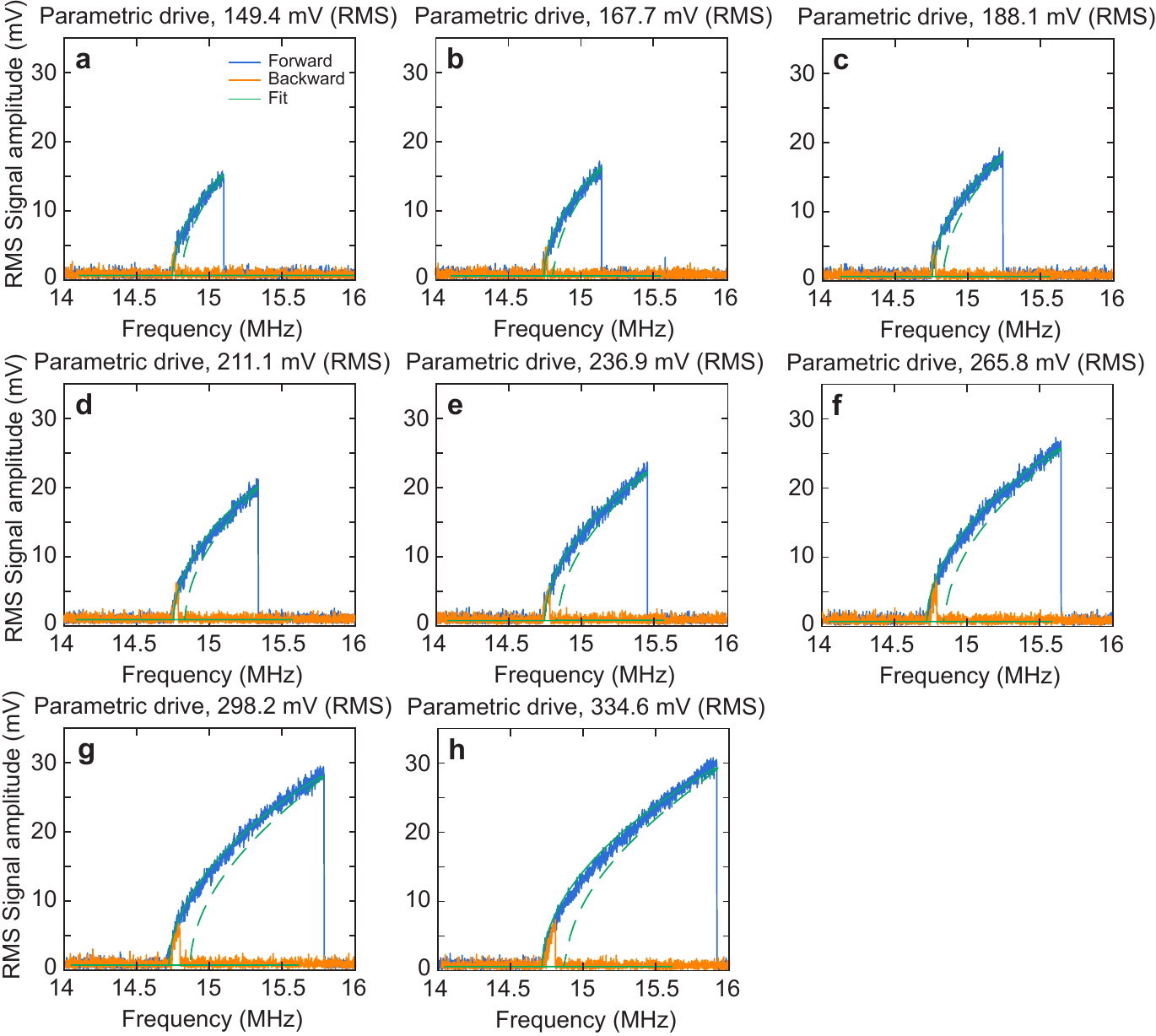}
\caption{Analysis of parametrically driven response of the fundamental mode of drum 1 (Figs. 2\textbf{a}, 3, 4\textbf{f} in the main text). The horizontal axis indicates the frequency at the analyzer port of the VNA, the frequency at the actuation port was doubled.  \label{fig:s3}}
\end{figure}
\begin{table}[h!]
\centering
\caption{Values obtained from the fits on the response from drum 1 in Figs. \ref{fig:s2}, \ref{fig:s3} and Figs. 2\textbf{a}, 3, 4\textbf{f}  in the main text.}
\label{tab:1l}
\begin{tabular}{llllllllll}
\hline
\hline\\
~ &    Direct drive    &    &       &                                                      & Parametric drive &        &    &       &                                                           \\\hline\\
RMS Drive (mV)       & $\mu$     &  $\nu$ & $\gamma$ & \begin{tabular}[c]{@{}l@{}}F $\times$ 10$^{-5}$\end{tabular}           & $\mu$     & $\nu$ & $\gamma$ & \begin{tabular}[c]{@{}l@{}}$\delta$ $\times$ 10$^{-2}$\end{tabular} \\ \hline
149.4          & 0.0030 & 36 & 250   & 1.42                                                                 & 0.0030 & 36 & 250   & 0.74                                                      \\
167.7          & 0.0030 & 37 & 245   & 1.6                                                                & 0.0030 & 36 & 220   & 1.01                                                      \\ 
188.1           & 0.0030 & 37 & 245   & 2.0                                                             & 0.0030 & 34 & 225   & 1.18                                                    \\ 
211.1           & 0.0030 & 37 & 245   & 2.5                                                           & 0.0030 & 34 & 225   & 1.31                                                       \\
236.9          & 0.0030 & 37 & 250  & 2.8                                                             & 0.0030 & 33 & 225   & 1.46                                                      
\\
265.8          & 0.0030 & 36 & 250   & 3.3                                                             & 0.0030 & 34 & 225   & 1.81                                                      
\\
298.2          & 0.0030 & 35 & 250   & 3.9                                                             & 0.0030 & 35 & 225   & 2.05                                                      
\\
334.6          & 0.0030 & 35 & 250   & 4.5                                                             & 0.0030 & 35 & 225   & 2.25                                                      
\\ \hline \hline
\end{tabular}
\end{table}

\pagebreak 
~
\pagebreak
\section*{S2: Additional experiment showing two stable phases}
\begin{figure}[h!]
\includegraphics{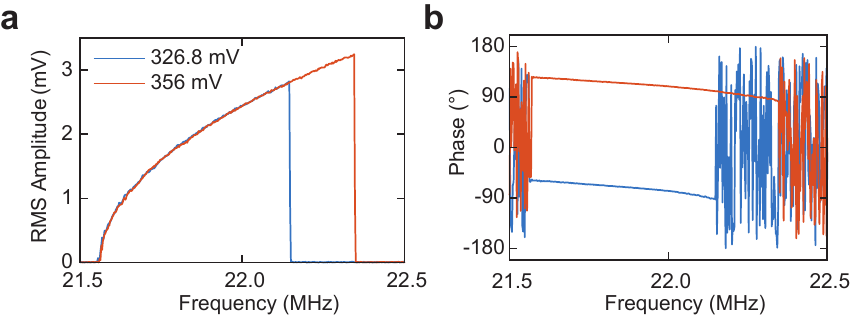}
\caption{Additional experiment of parametric resonance. Parametric excitation was achieved using the frequency doubler (as shown in Fig. 1 in the main text) instead of the frequency conversion on the vector network analyzer. (a) Amplitude of the response at two different driving powers and (b) the phase of these repsonses. The phase shows two stable phases separated by 180 degrees as expected. \label{fig:s4}}
\end{figure}
The frequency conversion option on the vector network analyzer loses information on the phase at which the resonator is oscillating. To show that the parametrically excited resonance has two stable phases separated by 180 degrees, the experiment was repeated by using the frequency doubler in the circuit used for the parametric amplification experiment (Fig. 1(c) in the main text). Using this, the VNA does not require to perform a frequency conversion and phase information is preserved. This results in the mechanical responses shown in Fig. \ref{fig:s4}.

\section*{S3: Additional discussion: mechanism for direct and parametric driving}
\begin{figure}[h!]
\includegraphics{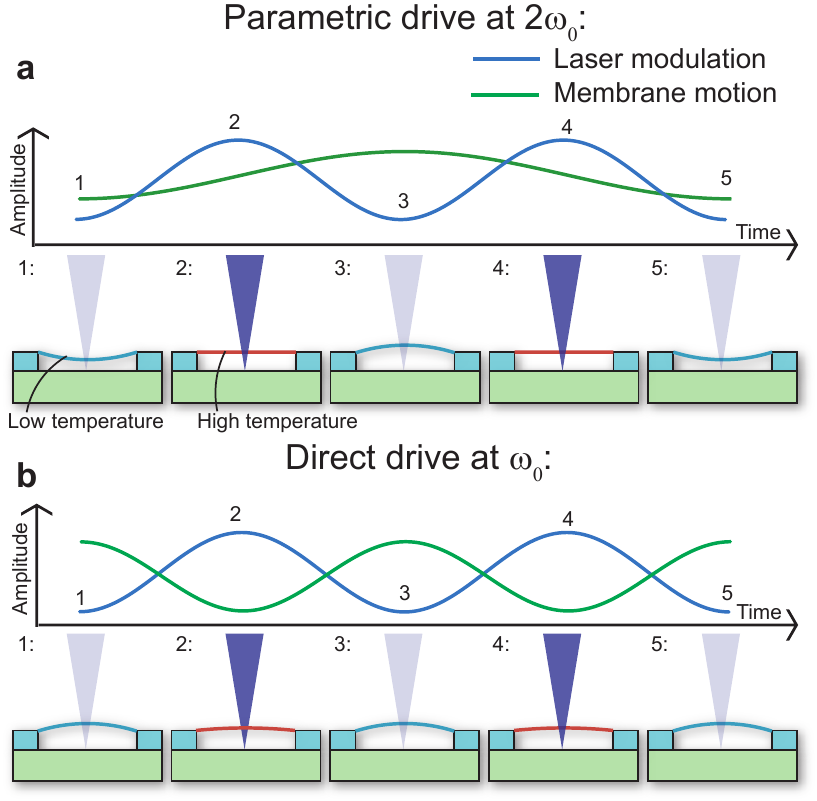}
\caption{Explanation of the actuation mechanisms of the opto-thermal drive. For illustration of the mechanism, it is assumed the membrane motion follows the force adiabatically (phase delays are omitted). A blue membrane represents low temperature and a red membrane represents high temperature. a) Parametric excitation, this is due to the pre-tension modulation of the membrane. Each time the tension is maximum the membrane passes through its equilibrium position, leading to a period doubling. This mechanism activates the resonance if the driving frequency is twice the resonance frequency. b) Direct excitation, which exists due to a small initial deviation from equilibrium. This mechanism does not cause period doubling, but instead it activates the resonance if the driving frequency is equal to the resonance frequency. \label{fig:3}}
\end{figure}
Opto-thermal driving leads to two mechanisms that can excite the resonance in the graphene resonators. Parametric drive (Fig. \ref{fig:3}a) occurs due to the modulation of pretension $n_0(t)$ in the membrane via laser heating and thermal expansion, since the stiffness term for the out-of-plane deflection field $w$ of the membrane determined by the pre-tension.
 Parametric driving will only activate the parametric resonance if the modulation of the blue laser is near twice the mechanical resonance frequency.

As demonstrated in the main text (Figs. 3, 4), the experiments also show a direct driving component. This can be explained \cite{aubin2004limit} by assuming a small initial membrane displacement $w_0$ from equilibrium (Fig. \ref{fig:3}b). In graphene resonators rippling, wall adhesion or out-of-plane crumples could lie at the root of such an initial displacement. 

 In order to analyze the data, we will derive the equations of motion (Eq. \ref{eq:EOM}) using a Lagrangian approach by including this initial deflection field. In this manner, the equations are reduced to a single-degree-of-freedom model that can be used to fit to the data, significantly simplifying the analysis. The derivation of this single degree of freedom model is shown below in section S4. 
   
\section*{S4: Equations of motion}
A Lagrangian approach is used for obtaining equations of motion of an optothermally excited monolayer graphene membrane. In this respect, the  potential energy of the thermally actuated circular membrane is obtained as \cite{marcoplates}:
\begin{equation}
U=\int_{0}^{2\pi}\int_{0}^{R} \frac{ h}{2} \Big(\sigma_{rr}(\epsilon_{rr}-\alpha \Delta T)+\sigma_{\theta \theta}(\epsilon_{\theta \theta}-\alpha \Delta T)+\tau_{r \theta} \gamma_{r \theta}\Big) r dr d\theta, \label{eq:potential energy}
\end{equation}
where $h$ is the thickness, $R$ is the radius, $\alpha$ is the thermal expansion coefficient, and $\Delta T$ is the temperature change in the membrane. Moreover, $\sigma_{rr}$ ,  $\sigma_{\theta \theta}$ ,  $\tau_{r \theta}$ ,  are the Kirchhoff stresses that can be obtained as follows:
\begin{equation}
\setlength{\jot}{10pt}
\begin{aligned}
&\sigma_ {rr}=\frac{E}{1-\nu^2}  (\epsilon_{rr}+\nu \epsilon_{\theta \theta}),\\
&\sigma _{\theta \theta}=\frac{E}{1-\nu^2}  (\epsilon_{\theta \theta}+\nu \epsilon_{rr}) \label{eq:Stress},\\
&\tau _{r \theta}=\frac{E}{2(1+\nu)} \gamma_{r \theta},
\end{aligned}
\end{equation}
in which $\epsilon_{rr}$, $\epsilon_{\theta \theta}$, and $\gamma_{r \theta}$  are the Green strains and are derived as:
\begin {equation}
\setlength{\jot}{10pt}
\begin{aligned}
&\epsilon_{rr}=\frac{\partial u}{\partial r}+\frac{1}{2}\Big(\frac{\partial w}{\partial r}\Big)^2+\Big(\frac{\partial w}{\partial r}\Big)\Big(\frac{\partial w_{0}}{\partial r}\Big), \\
&\epsilon_{\theta \theta}=\frac{\partial v}{r \partial \theta}+\frac{u}{r}+\frac{1}{2}\Big(\frac{\partial w}{r \partial \theta}\Big)^2+\Big(\frac{\partial w}{r \partial \theta}\Big)\Big(\frac{\partial w_{0}}{r \partial \theta}\Big),\\
&\gamma_{r \theta}=\frac{\partial v}{ \partial r}-\frac{v}{r}+\frac{\partial u}{r \partial \theta}+\Big(\frac{\partial w}{\partial r}\Big)\Big(\frac{\partial w}{r \partial \theta}\Big)+\Big(\frac{\partial w}{\partial r}\Big)\Big(\frac{\partial w_{0}}{r \partial \theta}\Big)+\Big(\frac{\partial w_{0}}{\partial r}\Big)\Big(\frac{\partial w}{r \partial \theta}\Big),
\end{aligned}
\end{equation}
where $u$, $v$ and $w$ are the radial,  tangential and transverse displacements, respectively. Moreover, $w_0$ is the deviation of the membrane from flat configuration, $E$ is the Young's modulus and $\nu$ is the Poisson's ratio.  

The temperature difference $\Delta T$ can be obtained by solving the following heat conduction equation:
\begin {align}
&\frac{\partial \Delta T}{\partial \tilde{t}}+\frac{\Delta T}{\tau}=\frac{P_{abs} \cos  (\omega \tilde{t} \,)}{C_{t}},
\end{align}
in which  $P_{abs}$ is the power absorbed by the membrane, $\tau$ is the thermal time constant \cite{dolleman2017optomechanics}, $C_{t}$ is the thermal capacitance, and $\tilde{t}$ represents the time variable.
\par For a membrane with fixed edges $u$ and $w$ shall vanish at $r=R$. Moreover, $u$ 
should be zero at $r=0$ for continuity and symmetry. Furthermore, assuming only axisymmetric vibrations ($v=0$ and $\partial{u}/\partial{\theta}=\partial{v}/\partial{\theta}=\partial{w}/\partial{\theta}=0$), the solution can be approximated as \cite{davidovikj2017young}:
\begin{align}
&w=x(\tilde{t}\,) J_{0} \Big(\alpha_{0}\frac{r}{R}\Big), \label{eq:lateral displacement} 
\end{align}
\begin{align}
&u= {u_{0} r}+ r (R-r) \sum_{k=1}^{\bar{N}} q_{k}(\tilde{t}) r^{k-1}. \label{eq:radial displacement}
\end{align}
Here it should be noted that for axisymmetric vibrations the shear strain $\gamma_{r \theta}$ would become zero. In equation (\ref{eq:lateral displacement}), {$x(\tilde{t})$} is the generalized coordinate associated with the fundamental mode of vibration. Furthermore, in equation (\ref{eq:radial displacement}), ${q_{k}(\tilde{t})}$'s are the generalized coordinates associated with the radial motion. Moreover, $J_{0}$ is the zeroth order Bessel function of the first kind and $\alpha_{0}=2.40483$. In addition, $\bar{N}$ is the number of necessary terms in the expansion of radial displacement and $u_{0}$ is the initial displacement due to pre-tension $n_{0}$ that is obtained from the initial stress $\sigma_0=n_{0}/h$ as follows :
\begin {equation}
u_{0}=\frac {\sigma_{0} (1-\nu)}{E}.
\end{equation}
The kinetic energy of the membrane neglecting in-plane inertia, is given by:
\begin{equation}
T=\frac{1}{2}\rho h\int_{0}^{2\pi} \int_{0}^{R} \Big( \frac{\partial w}{\partial \tilde{t}} \Big)^2 r dr d\theta. \label{eq:kinetic energy}
\end{equation}
The Lagrange equations of motion are given by:
\begin{equation}
\frac {d}{dt} \bigg(\frac{\partial T}{\partial \dot{\textbf{q}}}\bigg)-\frac{\partial T}{\partial \textbf{q}}+\frac{\partial U}{\partial \textbf{q}}=0, \label{eq:lagrange equations}
\end{equation}
and \textbf{q}=[${x(\tilde{t})}$,{$q_k$(\,$\tilde{t}$)}], $k=1,\ldots, \bar{N}$ is the vector containing all the generalized coordinates. Equation (\ref{eq:lagrange equations})  leads to a system of nonlinear equations comprising of a single differential equation associated with the generalized coordinate ${x(\tilde{t})}$ and $\bar{N}$ algebraic equations in terms of $q_{k}$($\tilde{t}$) . By solving the $\bar{N}$  algebraic equations it is possible to determine $q_k (\tilde{t})$ in terms of ${x(\tilde{t})}$ \cite{davidovikj2017young}. This will reduce the $\bar{N}$+1 set of nonlinear equations to the following Duffing-Matthieu-Hill equation:
\begin{equation}
m\ddot{x} + c_1 \dot{x} + c_2 x^2 \dot{x} + [k_1 + F_p \cos(\omega \tilde{t}\,)]x + k_2 x^2 + k_3 x^3 = F_d \cos(\omega \tilde{t}\,), \label{eq:EOM}
\end{equation}
where $\dot{(\bullet)}$ represents derivative with respect to time $\tilde{t}$ and  $m$ is the mass. $c_1$ and $c_2$ are the linear viscous damping coefficient and nonlinear material damping coefficient, respectively \cite{RN153,lifshitz2008nonlinear}. They are added to the equation of motion explicitly to introduce dissipation. $k_1$ represents the linear stiffness term dominated by the pre-tension $n_0$ and $F_p$ is the amplitude of parametric drive resulting from temperature variation $\Delta T$. Moreover, $k_2$ represents the quadratic non-linear stiffness coefficient due to imperfection $w_0$ and $k_3$ denotes the cubic non-linear stiffness coefficient arising from geometric nonlinearity. Finally, $F_d$ is the amplitude of direct drive term due to the presence of imperfection $w_0$, and $\omega$ is the excitation frequency. Indeed for a flat membrane, $k_2=F_d=0$. 

\section*{S5: Numerical simulations}
In order to perform the numerical simulations, equation (\ref{eq:EOM}) is normalized with respect to the mass ${m}$ of the membrane and the fundamental frequency ($t=\tilde{t}\omega_0$) as follows:
\begin{equation}
\ddot{x} + \mu \dot{x} + \nu x^2 \dot{x} + [\beta + \delta \cos(\Omega t)]x +\gamma_2 x^2+ \gamma_3 x^3 = F \cos(\Omega t). \label{eq:EOM dimensionless1}
\end{equation}
 Introducing an effective stiffness nonlinearity $\gamma$, whose value is given by $\gamma$=$\bigg(\gamma_3-\frac{10 \gamma_2^2}{9}$\bigg) \cite{nayfehosc}, equation (\ref{eq:EOM dimensionless1}) is reduced to:
\begin{equation}
\ddot{x} + \mu \dot{x} + \nu x^2 \dot{x} + [\beta + \delta \cos(\Omega t)]x + \gamma x^3 = F \cos(\Omega t), \label{eq:EOM dimensionless}
\end{equation}
where the normalized coefficients are given in table \ref{table:parameters}.

\begin{table}[h!t]
	\begin{center}
		\begin{tabular}{   >{$}c<{$} c  } 
			\hline
			\mathrm{Definition} & Normalized parameter\\
			\hline \\
			
			\dot{(\bullet)}=\frac{d(\bullet)}{dt} & Scaled time derivative\\
			\\
			\Omega=\frac{\omega}{\omega_0} & Non-dimensional excitation frequency \\
			\\
			\mu = \frac{c_1}{2 m \omega_0} & Scaled linear damping coefficient\\
			\\
			\nu = \frac{c_2 }{m \omega_0} & Scaled nonlinear damping coefficient\\
			\\
			\beta = \frac{k_1}{m \omega_0^2} & Scaled linear stiffness coefficient \\
			\\
			\delta= \frac{ F_p}{m \omega_0^2} & Scaled parametric excitation amplitude\\
			\\
			\gamma_2= \frac{k_2 }{m \omega_0^2} & Scaled nonlinear quadratic stiffness coefficient\\
			\\
			\gamma_3= \frac{k_3 }{m \omega_0^2} & Scaled nonlinear cubic stiffness coefficient\\
			\\
			\gamma= \gamma_3-\frac{10 \gamma_2^2}{9} & Scaled effective nonlinear stiffness coefficient\\
			\\
			F = \frac{ F_d}{m \omega_0^2} & Scaled direct excitation  amplitude \\
			\\
			\hline
		\end{tabular}
		\caption{Normalized parameter definitions}
		\label{table:parameters}
	\end{center}
\end{table}
Here it should be noted that, mass $m$ of the single layer graphene membrane is unknown. Without the exact mass value, optical transduction factors present between the voltage signal measured by the VNA during the experiment and the actual motion of the membrane in physical units cannot be calibrated. Thus, the  normalized coefficients shown in table \ref{table:parameters} include a linear transduction factor '$\kappa$' for the oscillation amplitude ($x=\kappa V_{1}$), $\eta$ for the parametric drive amplitude ($F_{p}=\eta V_{2}$) and $\lambda$ for the direct drive amplitude ($F_{d}= \lambda V_{3}$). Where ${V_1}, V_2$ and $V_3$ are voltage signals measured in the experiment. 

Finally, the equation (\ref{eq:EOM dimensionless}) is simulated using a pseudo arc length continuation and collocation technique \cite{AUTO} to detect bifurcations and obtain periodic solutions. The simulations are performed as follows:
\begin{enumerate}
\item The bifurcation analysis is carried out with the coefficient $F$ as the first continuation parameter and is incremented to the desired value in order to match the experimental direct response.
\item Once the desired value of $F$ is obtained, the parametric drive amplitude $\delta$ is used as the second continuation parameter and a value is chosen to replicate the experimental parametric response.
\item After reaching the desired $\delta$ value, the analysis is continued with the frequency ratio $\Omega$ as the final continuation parameter. This value is spanned around the spectral neighborhood of $\Omega=1$ and $\Omega=2$ in order to obtain the direct and parametric response curves.
\end{enumerate}

\section*{S6: Mechanical loss tangent of graphene}
In ref. \cite{davidovikj2017young} it is shown that the Duffing term $\gamma$ is proportional to the Young's modulus $E$:
\begin{equation}
\gamma = C E,
\end{equation}
where $C$ is a constant. In case of material damping, a complex Young's modulus can be introduced: $E = E'+iE''$ and the nonlinear stifness term $\gamma x^3$ near the resonance frequency $\omega_0$, for $x=x_0 \mathrm{e}^{-i \omega_0 t}$ becomes:
\begin{equation}
C E x^3 = C E' x^3 + C E'' \frac{x^2}{\omega_0} \dot{x} = \gamma x^3 + \nu x^2 \dot{x}.
\end{equation}
From this equation it can be seen that the loss tangent $\tan \delta_l = E''/E'$ \cite{lakes} can be calculated by the ratio $\nu/\gamma$ if the resonator is vibrating near its resonance frequency:
\begin{equation}
\tan \delta_l = \frac{\nu}{\gamma}.
\end{equation}

\end{document}